\documentclass[5p,twocolumn,10pt,times]{elsarticle}

\usepackage{enumerate}
\usepackage{amsmath}
\usepackage{booktabs}
\usepackage{pgf}
\usepackage{graphicx}
\usepackage{url}
\usepackage{hyperref}
\begin{document}

\begin{frontmatter}

\title{An Automated Approach for the Discovery of Interoperability}

\author{Duygu Sap}
\cortext[mycorresponingauthor]{Corresponding author}
\ead{duygusap@icsi.berkeley.edu}
\address{International Computer Science Institute, Berkeley, CA, USA.}

%\fntext[ICSI]{Since 1880.}

\author{Daniel P. Szabo}
\address{University of Wisconsin-Madison, WI, USA}



\begin{abstract}
\noindent In this article, we present an automated approach that would test for and discover the interoperability of CAD systems based on the approximately-invariant shape properties of their models. We further show that exchanging models in standard format does not guarantee the preservation of shape properties.\\
Our analysis is based on utilizing queries in deriving the shape properties and constructing the proxy models of the given CAD models\cite{on}. We generate template files to accommodate the information necessary for the property computations and proxy model constructions, and implement an interoperability discovery program called DTest to execute the interoperability testing.\\
We posit that our method could be extended to interoperability testing on CAD-to-CAE and/or CAD-to-CAM interactions by modifying the set of property checks and providing the additional requirements that may emerge in CAE or CAM applications.
\end{abstract}


\end{frontmatter}

%\linenumbers

\section{Introduction}
\subsection{Motivation}
\noindent Interoperability has been a challenging unsolved problem that relies on manual, error-prone solutions and costs billions of dollars annually \cite{geomintQ, geomintM}. 
Semi-automated verification of interoperability can be achieved by a set of limited tools. However, there does not exist any automated tools for the verification and the validation of interoperability solutions. This work may enable the next generation of automatically composable and reconfigurable systems, and support formal verification of the currently used standards. In this article, we focus on the theoretical framework we built in \cite{on}, 
and construct an algorithmic framework that can be used to apply the theory presented in \cite{on}. We also provide practical applications using the automated system we built based on the algorithmic framework we present here.\\
%\paragraph{Installation} If the document class \emph{elsarticle} is not available on your computer, you can download and install the system package \emph{texlive-publishers} (Linux) or install the \LaTeX\ package \emph{elsarticle} using the package manager of your \TeX\ installation, which is typically \TeX\ Live or Mik\TeX.
%
%\paragraph{Usage} Once the package is properly installed, you can use the document class \emph{elsarticle} to create a manuscript. Please make sure that your manuscript follows the guidelines in the Guide for Authors of the relevant journal. It is not necessary to typeset your manuscript in exactly the same way as an article, unless you are submitting to a camera-ready copy (CRC) journal.
%
%\paragraph{Functionality} The Elsevier article class is based on the standard article class and supports almost all of the functionality of that class. In addition, it features commands and options to format the
%\begin{itemize}
%\item document style
%\item baselineskip
%\item front matter
%\item keywords and MSC codes
%\item theorems, definitions and proofs
%\item lables of enumerations
%\item citation style and labeling.
%\end{itemize}
To our knowledge, there does not exist any work in the literature which has developed an algorithmic framework or an automated system that is capable of testing for the interoperability of CAD systems based on the interchangeability of their models with respect to their shape properties. By constructing such a framework and a system, we aim to show that it is possible to discover the interoperability between CAD systems with a pre-determined tolerance without translating formats or converting representations. We note that the interoperability we test for hinges on the interchangeability of a pair of CAD models that may be constructed by different CAD systems. Therefore, it may be identified as \textit{conditional interoperability}.\\ 
We further test if exchanging models in standard formats guarantees the preservation of the shape properties. We show that model transfers may result in unpredictable changes in the shape properties even when the standard formats are used.\\
\noindent The diagram below illustrates the logical structure of the automated system with its components. \\
%\begin{figure}[tbh!]
%    \centering
 % \resizebox{\linewidth}{!}{
%\smartdiagram[flow diagram:horizontal]{
 % User Interface, CAD Systems, Interoperability Diagnosis, Report} }
%\end{figure}

\begin{figure}[tbh!]
\centering 
\includegraphics[width=0.5\textwidth]{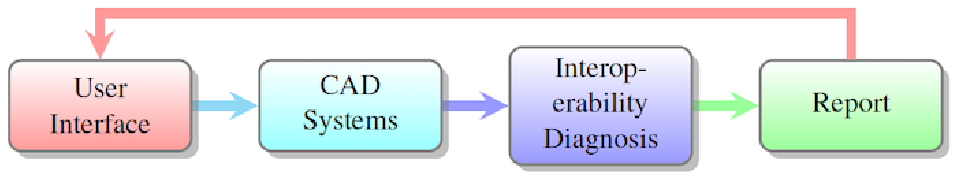}
%\caption{Automated testing and discovery for interoperability diagram}
%\label{fig:fig1}
\end{figure}

\noindent The interoperability diagnosis focuses on the interchangeability of CAD models with respect to a specified shape property with a given accuracy. Within the scope of this paper, we consider the following shape properties: geometric (convexity, centroid, Hausdorff distance), topological (homotopy types, manifoldness), integral (volume, surface  area), combinatorial (Euler characteristic).\\ 
\noindent We utilize the queries: PMQ, distance, integral, and we derive a result of the following form:\\
 `'CAD systems $C_1$ and $C_2$ that provide the respective models $M_1$ and $M_2$ can interoperate in carrying out a task that allows using $M_1$ and $M_2$ interchangeably with the given accuracy $\epsilon$ for the specified property`'.\\
\noindent We note that the differential properties could easily be added to the list of properties if the CAD software is capable of providing differential information through queries. Differential information could also be approximated via the lower level queries \cite{geomintQ}.

\subsection{Challenges}\label{12}
\noindent Here is the list of challenges we address in this article:
\begin{enumerate}
\item Investigating the automated verification of interchangeability of CAD models, thus, the automated verification of the conditional interoperability of CAD systems.
\item Validating the interoperability of CAD systems based on model transfer via direct data translation.% and/or queries. 
\item Investigating the invariant properties under data transfers and translations through Round-Robin testing.  
\end{enumerate}
Challenge $(3)$ may be utilized in solving the problem defined as the \textit{characterization of interoperability} in \cite{morad} in an automated manner.\\ 

\subsection{Related Work}
\noindent Interoperability is supported by semi-automated heuristic tools, requiring expertise and significant manual labor. 
\noindent There are various software companies that offer interoperability, translation verification and validation solutions.\cite{cadiq},\cite{cadfix},\cite{elysium},\cite{cadinterop} are some of the popular software tools provided by these companies. These software tools run checks on the models transferred and then utilize some automated or manual healing to compensate for the data and model quality loss. Thus, the models are modified to establish interoperability without any quantifiable measure.\\
\noindent Geometry validation testing based on a standard format, namely, STEP, has been carried out by CAx Implementor Forum since 1998\cite{cax}. Their methods are based on validating the shapes with respect to the properties such as centroid, volume and surface area for solid models, centroid and surface area for surface models, and centroid and total length for independent curves\cite{cax}. They also list a cloud of points method as an additional validation technique, but they do not provide a fully-developed and a practical methodology for this method. Their general testing procedure is based on the comparison of the results they derive from experimenting over various sets of models every six months. However, it is unclear why these sets of models are considered to constitute sufficient testing domains for validation since they do not seem to form bases for any geometric model space. \\
3D model validation techniques are also developed as military standards in \cite{31a}. Department of Defense (DoD) requires approved validation processes to show that 3D models are suitable for reference data. Since models may contain subtle defects that can prevent them from being used by downstream applications such as numerically-controlled manufacturing, finite element analysis, and inspection with coordinate-measuring devices, formal algorithmic validation processes are needed. The validation in \cite{31a} is done through identifying and classifying the defects in $3$D digital models along with their effects on various applications that use these models, and providing recommended tolerances and acceptance criteria for these models. However, the tolerances listed for the validation of different geometric components have some inconsistencies, and there is redundancy in the testing procedures. A thorough analysis of these military standards is provided in \cite{31d}.
\subsection{Contributions}
\noindent The main contributions of this article can be listed as follows:
\begin{itemize}
    \item The first known theoretically supported automated system for the interoperability of CAD systems with respect to shape properties, namely, geometric, topological and integral properties. The automated system presented here trivializes the construction of a system with an extended set of properties such as the physical and material properties.
   % \item A practical theoretical methodology based on proxy models serving as shape identifiers. 
    \item The first known algorithmic framework that could be used for verifying CAD model interchangeability, and validating CAD model transfers or format translations via queries.
 
    \item A method for investigating the properties preserved under model transfers via standard formats, which in turn would indicate the ability to predict and maintain the model quality for long-term archival and retrieval (LOTAR) \cite{lotar}. 
    \item Demonstrated applicability of the theory developed in \cite{on} to current practices that are being tested by NIST \cite{cax}.

\end{itemize}

\section{Research Methodology}\label{sec2}
\noindent Our methodology is based on building proxy models that would substitute for the CAD models in property-based comparisons and investigating shape equivalences with predictable accuracies to determine the interchangeability of given CAD models, or validate a CAD model translation\cite{on}. Proxy model and shape equivalence constructions are based on a parameter $\epsilon$ which is determined by the accuracy of the point membership classification query (PMQ), tolerances and algorithm precisions of systems and the minimum feature sizes of the CAD models\cite{on}.\\
In the following subsections, we present an overview of our theoretical framework and provide details on the structures and inner mechanisms of our algorithmic framework and  automated approach.
\subsection{Problem Definitions}
\noindent We mainly consider two problems:
\begin{itemize}
    \item Automated Verification of Model Interchangeability
    \item Automated (Round-robin) Testing %for the Characterization of Interoperability
\end{itemize}
\subsubsection{Automated Verification of Interchangeability}
%\noindent \textit{Statement:} 
\noindent Given two CAD models, determine if the models are interchangeable based on a shape property with the stated accuracy
\subsubsection{Automated (Round-robin) Testing}
%\noindent \textit{Statement:} 
\noindent Given a CAD model, read and write the model in STEP format multiple times in the same system or circulate it over different systems and observe the variations in the properties to determine if or when any shape property value deviation converges to zero.  
\subsection{Theoretical Framework}
\noindent Our theoretical framework for the verification of interoperability (detailed description of which can be found in \cite{on}) enables determining the interchangeable usability of the CAD models created by distinct CAD systems through a query-based data analysis. The query-based approach allows us to test and discover the interoperability of CAD systems and model representations via a set of queries instead of translators and/or transferring files. Queries are computable functions with semantics specified with respect to a standard reference. In query-based interoperability scenarios, systems are allowed to have different representations and algorithms, and they retain their copies of the models separately\cite{geomintQ, morad}.\\
In \cite{on}, we provided the sufficient conditions for establishing a correspondence based on a topological equivalence and a geometric similarity between model instances authored by distinct CAD systems. We referred to the proxy model notion defined by Hoffmann et al in \cite{geomintQ} and described how to construct and utilize proxy models of given CAD models. 

A proxy model
\begin{itemize}
\item substitutes for the CAD model in property calculations,
\item may take different forms (e.g., a point cloud, union of balls, a graph, an algebraic complex such as a Cech complex),
\item may be built in one of the systems or exist as a separate (abstract) reference model,
\item could be set as one of the CAD models in a model comparison scenario,
\item provides estimates for the models' properties that depend on model-specific data, which is implicitly related to systems' attributes.

\end{itemize}
 
\noindent We note that the proxy model constructions and the property computations could be completely carried out through a query-based approach as a result of which the model instances are liberated from their system-dependent representations \cite{on}. 

\subsection{Algorithmic Framework}
\noindent In this section, we present the algorithmic framework that enables building proxy models of the given CAD models and computing the properties of these proxy models to allow a property-based model comparison. We run the tests over the template files that substitute for the models. Thus, the testing procedure does not require model transfers or translations, and the property information can be derived directly through  standardized sources. Figure~\ref{fig1} illustrates the structure of the algorithmic framework.\\
We note that this algorithmic framework could be used in executing tests on CAD models that are saved in different formats in distinct CAD systems as well as on CAD models given in standard formats.

\begin{figure}[tbh!]
\centering 
\includegraphics[width=0.5\textwidth]{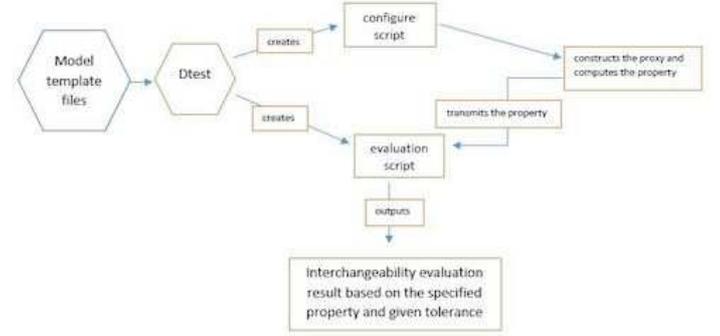}
\caption{Automated testing and discovery for interoperability diagram}
\label{fig:fig1}
\end{figure}
%\chapter{Technical Details}
\noindent Every CAD model either comes with a template file or in standard format, namely, STEP.
%\subsubsection{Algorithmic Framework Components}
\subsubsection{DTest}

\textit{DTest} is the key component of the algorithmic framework and the executable component of the automated system.\\
\noindent Here is the list of tasks \textit{DTest} carries out:
\begin{itemize}
\item Running on model template files and picking up the property with its tolerance entered by the user

\item Determining the set of query responses that should be picked from the template files
\item Collecting the necessary information from the template files to determine parameters such as the ball radius $\epsilon$ : $\epsilon_i+\alpha_i <\epsilon <\delta_i$, where $\alpha_i$ and $\epsilon_i$ are the respective algorithm precision and absolute tolerance of the CAD system $C_i$, and $\delta_i$ is the minimum feature size of the model $M_i$ created by $C_i$.
\item Creating the \textit{configure} for constructing a proxy and computing properties

\item Creating the \textit{evaluation} that would evaluate the results of the tests with respect to the allowable tolerance level specified by the user for the specified property  

\end{itemize}
%\end{itemize}
We note that if the CAD systems the CAD models are authored by use different scripting languages, wrapper functions \cite{occ} need to be used to build a standardized coding environment based on a single programming language.
%\color{black}
\subsubsection{Configure} 
\noindent \textit{Configure} is responsible for the following tasks:
%\begin{minipage}{30em}
\begin{itemize}
\item Constructing the proxy model that can substitute for $M_i$ in the property comparison. 
\item Computing the relevant property of the proxy models with the parameters it receives from the template files via \textit{DTest}
\end{itemize}
\noindent Note that the choice of the proxy depends on what is sufficient for computing the property. Moreover, in some cases, \textit{configure} does not need to define functions for computing model properties. For example, if we are investigating the similarity of models in the Hausdorff metric, then it computes the Hausdorff distance between the proxy models, which would be point clouds in such case.
\subsubsection{Evaluation}
\noindent \textit{Evaluation} carries out the following tasks:
\begin{itemize}
\item Compares the properties of the models
\item Derives and outputs an interoperability report with respect to the standards that \textit{DTest} sets for the model interchangeability based on the information provided by the user% and level of tolerance provided by the user for the specified property
\end{itemize}
\subsubsection{Template File}

\noindent Template files are model-specific, and they would ideally accommodate the system information that was active in the design environment where the model was constructed (See Appendix A). However, in this research we include the system information active in the environment where the model was read in STEP format in our experiments. It is important to note that the template files that do not provide the authoring system specifications are likely to suffer from the drawbacks of the existing standard formats. The current automated system requires the users to create the template files.\\
\noindent A template file for a model denoted by $M_i$ has the following content:
%\textbf{Template File for $M_i$}
\begin{itemize}
\item Authoring CAD system $C_i$ 
\item API options
\item Scripting languages
\item System tolerances: absolute tolerance $\epsilon_i$, angular tolerance $\epsilon_i^a$
\item Algorithm precisions: reading precision $\alpha_i^r$, writing precision $\alpha_i^w$, PMQ accuracy 
\item List of queries supported by $C_i$
\item Measurement units
%\item Query responses of $M_i$ 
\item Topological class of $M_i$
\item Minimum feature size $\delta_i$
%\item The coordinate information of the points that return ``in"/``on" to point-membership queries (saved as .TXT)  
\end{itemize}

\begin{figure}[tbh!]
    \centering
    \includegraphics[width=0.3\textwidth]{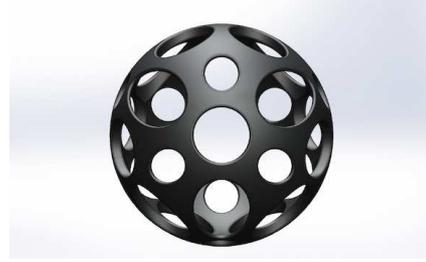}
    \caption{A CAD model provided as a STEP file by \cite{gameC}. For this model, the minimum feature size, $\delta_i$=min $\{r: r$ is the radius of a hole on the sphere$\}$}.
    \label{fig:ex}
\end{figure}

\noindent Note that we could split a template file into two sections  concerning the information content: system-specific section and model-specific section.

\section{Automated System Design}

\noindent In this section, we provide the automated testing procedure for the interoperability of the CAD systems. The testing is done over model template files by focusing on the invariance of a given shape property within a specified interval. For example, when we run \textit{DTest} on the model template files $Temp(M_1)$ and $Temp(M_2)$ by stating a shape property $P$ along with an accuracy $\epsilon$, we test for the interchangeability of $M_1$ and $M_2$ with respect to $P$ with $\epsilon$ accuracy. The test result would be positive if $|P_{M_1}-P_{M_2}|\le \epsilon$, thus, the models would be deemed interchangeable with $\epsilon$ accuracy for the applications that only use the property $P$. This further yields a conditional interoperability of the CAD systems $C_1$ and $C_2$ since for any application that needs to use only the property $P$ of the shape represented by $M_i$ in the CAD systems $C_i$, the systems $C_1$ and $C_2$ can interoperate with accuracy $\epsilon$ in using their models $M_1$ and $M_2$ interchangeably.

\noindent In the following subsections, we show how the automated system works for the two problems stated in Section \ref{sec2}. 
\subsection{Testing for Model Interchangeability}\label{31}
\noindent Here is the list of problems we addressed using the automated system to determine model interchangeability:
\begin{enumerate}[(i)]
    \item \textbf{Verification of Model Interchangeability:} Suppose $M_1$ and $M_2$ are CAD models created in the respective CAD systems $C_1$ and $C_2$. To decide on the interchangeability of $M_1$ and $M_2$, we construct their proxies $M^1$ and $M^2$ with the query responses and compute the properties of these proxies via \textit{configure}. Then,  \textit{evaluation} would return a quantitative comparison between the models of $M_1$ and $M_2$ by using the accuracy of $M^i$ in computing the properties of $M_i$ and the computed values passed on from \textit{configure}.
    \item \textbf{Translation Validation:} Suppose $M_1$ is a CAD model authored by a CAD system $C_1$ and let $M^t_1$ be its translated version in another CAD system $C_2$. To validate the translation, we follow the same strategy in Step-i by setting $M_2=M^t_1$.
\end{enumerate}
Automated testing for the interoperability of CAD systems based on the interchangeability of their models can be carried out by the following steps:
%These could be in an automated manner as follows:
\begin{itemize}
\item Fix a model instance $M_i$ in a CAD system $C_i$
\item Identify if the system $C_j$ would validate the shape properties of $M_i$ (interoperability of $C_j$ with $C_i$)
\item Repeat the same procedure by switching the roles of $C_i$ and $C_j$
\end{itemize} 

\subsection{Round-Robin Testing}\label{32}
\noindent CAD users may not know when and in which CAD system the STEP files they are working with were created and how many times they were transferred. To verify the reusability of CAD models after iterative transfers in a standard format, we investigate if any of the properties we check stabilizes at a particular iteration during Round-Robin tests\cite{wiki}.\\ 
\noindent Suppose $M_i$ is a CAD model created by a CAD system $C_i$ and saved as a STEP file. Let $\{M_{i_j}\}_{j=0}^k$ be the sequence of models in the Round-Robin testing process for $k$ rounds of tests.\\ We mainly consider two types of round-robin tests:
\begin{enumerate}[(i)]
    \item Read and write $M_i$ in $C_i$ $k$ times and compute the properties of $\{M_{i_j}\}$ to observe the changes and determine if there exists $l \in \{1,\dots,k\}$ such that $P(M_{i_l})=P(M_{i_{l-1}})$ where $P$ denotes a property function.
    \item Let $C_m$ be another CAD system and run the round tests between the systems $C_i$ and $C_m$, that is, $M_{i_j}$ is a STEP file generated by $C_i$ and $C_m$ for $j$ even and odd, respectively. The goal is the same as in the case $(i)$. 
\end{enumerate}
We note that in these experiments $C_i$ and $C_m$ denote OpenCASCADE and Rhinoceros, respectively.\\
The diagram along with the list of actions below shows how the invariance of shape properties is investigated. 
%
%\begin{figure}[tbh!]
%\centering
%    \resizebox{\linewidth}{!}{
%\smartdiagram[flow diagram:horizontal]{
 % CAD System $C_i$, CAD System $C_j$, CAD System $C_i$, Invariant Property} 
%}
%\end{figure}
\begin{figure}[tbh!]
\centering 
\includegraphics[width=0.5\textwidth]{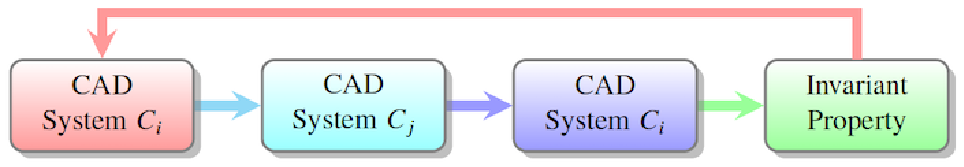}
%\caption{Automated testing and discovery for interoperability diagram}
%\label{fig:fig1}
\end{figure}

\begin{enumerate}
\item Construct a solid model in system $C_i$
\item Transfer the model to another system $C_j$, then transfer the translated model back to $C_i$
\item Investigate the shape properties that would change less than a specified value under the recursive format translations or model transfers
\begin{enumerate}[i.]
\item Determine the property with a negligible invariance at the end of a round test 
\item Repeat $(2)$ and $(i)$ 
\end{enumerate}
\end{enumerate}

\section{Experiments}
\noindent In this section, we illustrate some of the experiments we carried out following our automated approach. In Section~\ref{mi}, the models are represented by template files. In Section~\ref{rt}, the models are saved in STEP format, and they either maintained their format throughout the experimentation or a format translation took place before they are saved as STEP again.
\subsection{Testing for Model Interchangeability}\label{mi}
\noindent Here is how we execute \textit{DTest}:

\begin{verbatim}
./DTest <TemplateFile1> <TemplateFile2><TestName> 
\end{verbatim}
\begin{verbatim} 
 <Tolerance> 
\end{verbatim}
\noindent \textit{TestName} is chosen from the list $\{$volume, surface area, Hausdorff distance, centroid$\}$.\\

\noindent In the following experiment, we examine the interchangeability of a CAD model read by two distinct systems.

  \subsubsection{Experiment}
\begin{figure}
    \centering
    \includegraphics[width=0.4\textwidth]{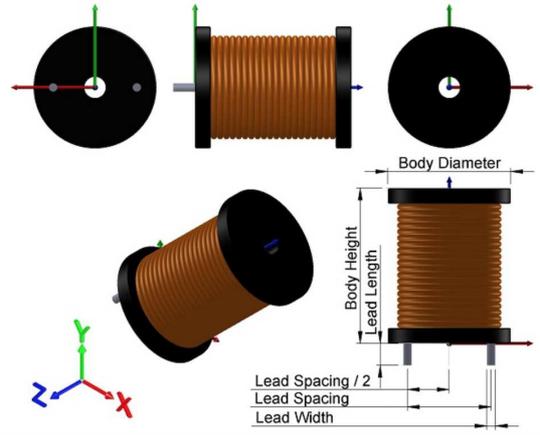}
    \caption{Electrical Coil \cite{pcb}.}
    \label{fig:coil}
\end{figure}
\noindent In this experiment, the coil model in Figure~\ref{fig:coil} is read as a STEP file by OpenCASCADE and converted to a 3DM format in Rhinoceros. We query the systems to build the template files, then run \textit{DTest}. The template file is provided in Appendix-A.
Table~\ref{tab:my_label} illustrates the key components of the template files.
\begin{table}[tbh!]
  \centering
 %\begin{center}
  \resizebox{0.4\textwidth}{!}{
  %\begin{minipage}{\textwidth}
    \begin{tabular}{c c c c c c  }
  % \centering
    \toprule
     CAD System   & OpenCASCADE & Rhinoceros& & &   \\
     \hline
     System Tolerance    & $1e-5$  & $1e-5$& & &  \\
     \hline
     Ball Radius & $\epsilon$& $\epsilon$ & & &   \\
     \hline
     PMQ accuracy & $2e-1$& $2e-1$& & &  \\
     \hline
     Model & Non-Convex & Non-Convex & & &  \\
     \bottomrule
    \end{tabular}

   %  \end{minipage}
     }
\caption{Main specifications for the electric coil model in the template file}
\label{tab:my_label}
    % \end{center}
\end{table}
%\newpage

\noindent Running DTest on (RhinoModel) Rcoil.xml and (OCCModel) Ocoil.xml Property $\epsilon=0.0001$\\
\noindent \underline{Output-1:}\\
Volume:
Systems Rhino and OpenCASCADE have incompatible volumes with a difference of 11.06102413\\
Volume of first proxy model: 476.73668518,\\ Volume of second proxy model: 487.79770932\\
Surface Area:
Systems Rhino and OpenCASCADE have incompatible areas with a difference of 132.73228961\\
Surface area of first proxy model: 5720.84022219,\\ Surface area of second proxy model: 5853.57251180\\
Hausdorff Distance:
Systems Rhino and OpenCASCADE have an incompatible Hausdorff Distance of 1.18016199\\ \\
\underline{Report}:\\
Rhinoceros and OpenCASCADE that provide the respective models, Rcoil and Ocoil, cannot interoperate in carrying out a task that allows using Rcoil and Ocoil interchangeably with the given accuracy $\epsilon$ for the specified property.\\

\noindent The following shows an output of the same experiment done with ball radius equal to $1e-1$.\\

\noindent Running DTest on (RhinoModel) Rcoil.xml and (OCCModel) Ocoil.xml Property $\epsilon=0.001$\\
\noindent \underline{Output-2:}\\
Volume:
Systems Rhino and OpenCASCADE have incompatible volumes with a difference of 1058.03869381\\
Volume of first proxy model: 4907.00478882,\\ Volume of second proxy model: 3848.96609501\\
Surface Area:
Systems Rhino and OpenCASCADE have incompatible areas with a difference of 3907.53289433\\
Surface area of first proxy model: 5823.00495112,\\ Surface area of second proxy model: 1915.47205679\\
Hausdorff Distance:
Systems Rhino and OpenCASCADE have an incompatible Hausdorff Distance of 4.44160210.\\
\underline{Report}:\\
Rhinoceros and OpenCASCADE that provide the respective models, Rcoil and Ocoil, cannot interoperate in carrying out a task that allows using Rcoil and Ocoil interchangeably with the given accuracy $\epsilon$ for the specified property.\\

\noindent The results emphasize the significance of the ball radius $\epsilon$ in the property comparison that is based on proxy models. For more reliable results and better approximations, the ball radius should be chosen closer to the system tolerance and non-convex shapes should be avoided in order to capture small features and get $\epsilon$ dependent bounds on the integral properties as the theory suggests \cite{on}.
 
\subsection{Round-robin Testing}\label{rt}
\noindent The experiments we present in this section show that the models could change slightly in an almost random way during transfers in STEP format.
\subsubsection{Experiment 1}
\noindent This experiment exemplifies the case $(i)$ of Section~\ref{32}. 
Our experiments on the STEP files provided by NIST \cite{cax} show that after a couple of round tests, the variations in the properties become hard to detect within available precision.

\begin{figure}[tbh!]
    \centering
    \includegraphics[width=0.3\textwidth]{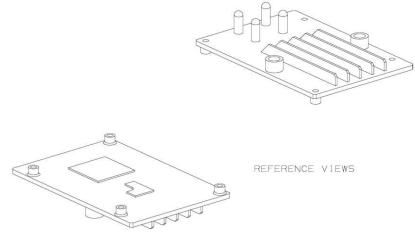}
    \caption{Test model labelled as NIST $904$. This model is chosen from the list of models provided for a testing-round carried out by CAx Implementor forum \cite{cax}}
    \label{fig:nist1}
\end{figure}
\noindent The Round-Robin tests of STEP files in OpenCASCADE show that the shape properties, which are considered as the basic properties to be checked for the interchangeable use of CAD models in engineering applications, exhibit a noticeable amount of change in the first few rounds (See Table~\ref{tab:mytab3} \& \ref{tab:mytab4}) and then they stabilize in the third or fourth transfer in general. We note that this deduction is based on the tests we ran over a large set of STEP files provided by \cite{cax} for a round test.\\
\noindent Additionally, the graph in Figure~\ref{fig:robinG} shows how reading and writing the same model as STEP in the same system multiple times can result in relatively random changes in properties such as volume and surface area.
\begin{table}[tbh!]
\centering % Centers the table on the page, comment out to left-justify
 \resizebox{0.5\textwidth}{!}{
\begin{tabular}{l c c c c c } 
\toprule % Top horizontal line
& \multicolumn{5}{c}{NISTModel.STEP Integral Property Computations} \\ % Amalgamating several columns into one cell is done using the \multicolumn command as seen on this line
\cmidrule(l){2-6} % Horizontal line spanning less than the full width of the table - you can add (r) or (l) just before the opening curly bracket to shorten the rule on the left or right side
Model & Volume  & Area &  & &  \\ % Column names row
\midrule % In-table horizontal line
$M_i$ & 22869.801015681573 & 20900.779662695128 &  & &\\ % Content row 1
$M_{i_1}$& 22869.801015681736 & 20900.779662695266 &  & & \\ % Content row 2
$M_{i_2}$ & 22869.80101568175 & 22869.80101568175 & & &  \\ % Content row 3
$M_{i_3}$ & 22869.80101568175 & 22869.80101568175 &  & &  \\ % Content row 4
 % &   &   &  &  & \\ % Content row 5
\midrule % In-table horizontal line
\midrule % In-table horizontal line
Stabilized in & Round 3 &  & &  \\ % Summary/total row
\bottomrule % Bottom horizontal line
\end{tabular}}
\smallskip 
\caption{Round-robin Test Integral Property Check Results. The model NIST 904 is read by OpenCASCADE.} % Table caption, can be commented out if no caption is required
\label{tab:mytab3} % A label for referencing this table elsewhere, references are used in text as \ref{label}
\end{table}
\begin{table}[tbh!]
  \centering
  \resizebox{0.5\textwidth}{!}{
  %\begin{minipage}{\textwidth}
% Add the following just after the closing bracket on this line to specify a position for the table on the page: [h], [t], [b] or [p] - these mean: here, top, bottom and on a separate page, respectively
%\centering % Centers the table on the page, comment out to left-justify
%\scalebox{1.05}{
\begin{tabular}{l l l l l l p{5cm}}% The final bracket specifies the number of columns in the table along with left and right borders which are specified using vertical bars (|); each column can be left, right or center-justified using l, r or c. To specify a precise width, use p{width}, e.g. p{5cm}
\toprule % Top horizontal line
& \multicolumn{5}{l}{NISTModel.STEP Geometric Property Computation} \\ % Amalgamating several columns into one cell is done using the \multicolumn command as seen on this line
\midrule
%\cmidrule(r){2-6} % Horizontal line spanning less than the full width of the table - you can add (r) or (l) just before the opening curly bracket to shorten the rule on the left or right side
Model &  Centroid  coordinates & & & \\ % Column names row
\midrule % In-table horizontal line
$M_i$ & (51.67985064942907,  34.72604686416809, & & &\\ &  2.0243846113346495) & & &   \\ % Content row 1
$M_{i_1}$ & (51.679850649429625,  34.7260468641684, & & &\\& 2.0243846113346287) & & &\\ % Content row 2
$M_{i_2}$ & (51.679850649429625,  34.72604686416839, & & &\\ & 2.0243846113346287) & & & \\ % Content row 3
$M_{i_3}$ & (51.679850649429625,  34.72604686416839, & & &\\ & 2.0243846113346287)& & &  \\ % Content row 4
%$M_{i_4}$ & 0.916  \\ % Content row 5
\midrule % In-table horizontal line
\midrule % In-table horizontal line
Stabilized in  & Round 3 & & &\\ % Summary/total row
\bottomrule % Bottom horizontal line
\end{tabular}
%\end{minipage}
}
%}
\smallskip 
\caption{Round-robin Test Geometric Property Check Results. The model NIST $904$ is read by OpenCASCADE.} % Table caption, can be commented out if no caption is required
\label{tab:mytab4} % A label for referencing this table elsewhere, references are used in text as \ref{label}
\end{table}
\begin{figure}[tbh!]
       \centering
       \includegraphics[width=0.35\textwidth]{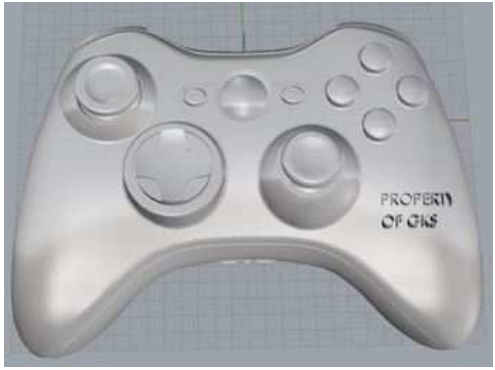}
       \caption{Game Controller \cite{gameC}.}
      \label{fig:game}
   \end{figure}

   \begin{figure}[tbh!]
       \centering
       \includegraphics[width=0.5\textwidth]{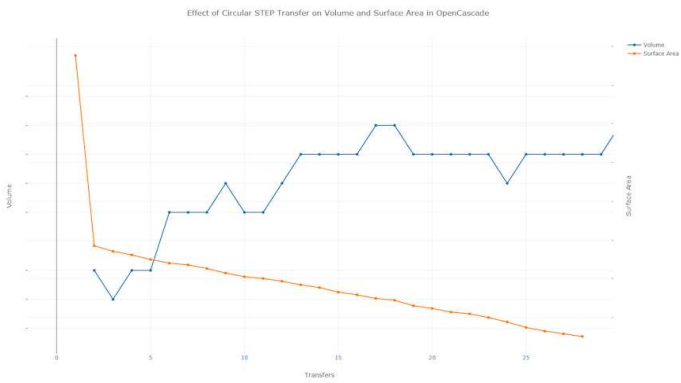}
       \caption{Volume and surface area changes for the Game Controller model in Figure~\ref{fig:game} during Round-robin transfers in OpenCASCADE}
       \label{fig: robinG}
   \end{figure}   
%\newpage

 \begin{table}[tbh!] % Add the following just af%ter the closing bracket on this line to specify a position for the table on the page: [h], [t], [b] or [p] - these mean: here, top, bottom and on a separate page, respectively
\centering % Centers the table on the page, comment out to left-justify
  \resizebox{0.5\textwidth}{!}{
\begin{tabular}{l c c c c c } % The final bracket specifies the number of columns in the table along with left and right borders which are specified using vertical bars (|); each column can be left, right or center-justified using l, r or c. To specify a precise width, use p{width}, e.g. p{5cm}
\toprule % Top horizontal line
& \multicolumn{5}{c}{GameController.STEP Integral Property Computations} \\ % Amalgamating several columns into one cell is done using the \multicolumn command as seen on this line
\cmidrule(l){2-6} % Horizontal line spanning less than the full width of the table - you can add (r) or (l) just before the opening curly bracket to shorten the rule on the left or right side
Model & Volume  & Area &  & &  \\ % Column names row
\midrule % In-table horizontal line
$M_i$ & 42744.69330223343 &365837.36461902654 &  & &\\ % Content row 1
$M_{i_1}$& 42744.68897336077 & 365840.19308225340 &  & & \\ % Content row 2
$M_{i_2}$ & 42744.68897338038 & 365840.19308251600 & & &  \\ % Content row 3
$M_{i_3}$ & 42744.68897338038 & 365840.19308251573 &  & &  \\ % Content row 4
$M_{i_4}$ &  42744.68897338038 & 365840.1930825155  &  &  & \\ % Content row 5
\midrule % In-table horizontal line
\midrule % In-table horizontal line
Stabilized in & Round 3 & +10 & &  \\ % Summary/total row
\bottomrule % Bottom horizontal line
\end{tabular}}
\smallskip 
\caption{Round-robin Test Integral Property Check Results. The model shown in Figure~\ref{fig:game} is read by OpenCASCADE.} % Table caption, can be commented out if no caption is required
\label{tab:tab6} % A label for referencing this table elsewhere, references are used in text as \ref{label}
\end{table}

\begin{figure}[tbh!]
       \centering
       \includegraphics[width=0.4\textwidth]{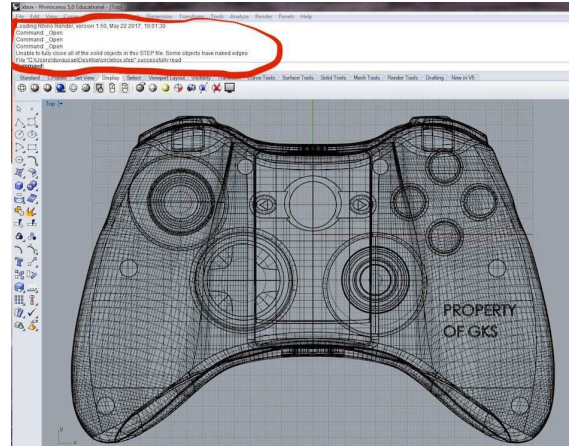}
       \caption{Game Controller model in Figure~\ref{fig:game} after it is read by Rhino as a STEP file}
      \label{fig:gameR}
   \end{figure} 

   \begin{figure}[tbh!]
       \centering
       \includegraphics[width=0.4\textwidth]{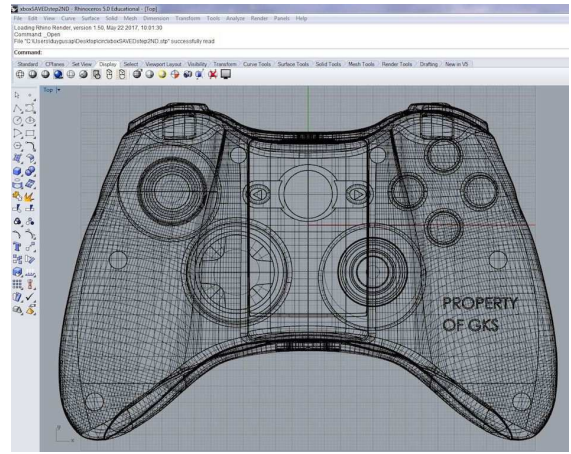}
       \caption{Game Controller model in Figure~\ref{fig:gameR} after it is saved as 3DM and STEP by Rhino, respectively.}
      \label{fig:XBOXstep2}
   \end{figure}

%
%\newpage
\subsubsection{Experiment 2}
\noindent In this experiment, we aim to observe the changes due to format translations, namely, translating a model from a standard format to a native format and vice versa. In this respect, we read and rewrite the model illustrated in Figure~\ref{fig:game}. We read the model as a STEP file in Rhinoceros. Then, we translate it to the native format, 3DM, of Rhinoceros. After that, we translate it back to STEP, we observe that Rhinoceros was unable to fully close all of the solid objects in the STEP file and some objects had naked edges(See Figure~\ref{fig:gameR}).  
\subsubsection{Experiment 3}
\noindent This experiment exemplifies the case $(ii)$ stated in Section~\ref{32}. Here, we mainly test the STEP conversion capabilities of the systems and search for the data loss in terms of the properties.\\
We run Round-Robin tests between a pair of distinct CAD systems, namely, OpenCASCADE and Rhinoceros. \\
During the Round-robin testing of STEP files in the following sequence of CAD systems:  \begin{verbatim}   OCC--(STEP)-->Rhino--(STEP)-->OCC\end{verbatim} where OCC denotes OpenCASCADE, and Rhino stands for Rhinoceros, we observe the following:

\begin{itemize}
\item \textbf{Change in topology:} Rhinoceros is unable to fully close all of the solid objects in this STEP file. Some objects have naked edges.
\item \textbf{Change in geometry:} Shifts occur in the centroid positions 
\end{itemize}
\noindent These experiments show that there is a need for a mechanism that would provide measures on the quality of the standardization. The "standard" formats such as STEP  can only offer a weak standardization as a result of which additional healing or repair software accompanies the model transfers in standard formats. We also note that the limitation on the number of round testing for STEP file transfers poses a limitation on predicting the convergence properties of the data loss. As a result, the data loss during transfers has an unpredictable nature.
%\newpage

\section{Technical Challenges}
\noindent Here is a list of technical challenges we faced during this research:
\begin{enumerate}[(i)]
\item \textbf{Limited access to commercial systems:} 
This hinders the development of the testbed for Round-Robin experiments. %We focused on STEP-based interoperability practices as carried out by NIST.
%\begin{itemize}
 %   \item The CAD systems we had free access to were highly incompatible as mentioned in Section VI.
%\end{itemize}
\item \textbf{Theoretical computational complexity:} 
Complete verification with respect to an external proxy model may become intractable in the presence of small features or high precision. However, this challenge can be alleviated by using selective testing, localization and statistical measures.
\item \textbf{Operating System \& CAD software incompatibility:} Some CAD software do not work on every operating system. For example, there is no downloadable version of Rhinoceros for Linux.
\item \textbf{Programming language incompatibility:} Different scripting languages may be used by different CAD software, which was the case for the CAD software used in this project. Rhinoceros uses IronPython, which is firmly integrated with the .NET Framework\cite{iron}, whereas the OpenCASCADE community uses OCE, which is a C++ 3D modeling library\cite{oce}.
\end{enumerate}

\section{Conclusion and Future Direction}
\noindent In this article, we presented an algorithmic framework and an automated approach that would test for and discover conditional CAD-to-CAD interoperability which is based on the approximately-invariant shape properties between two given CAD models. We further showed that exchanging models in standard format does not guarantee the preservation of the shape properties.\\
\noindent We posit that our method could be extended to interoperability testing on CAD-to-CAE and/or CAD-to-CAM systems by modifying the set of property checks and testing the additional conditions that would be required by the CAE or CAM applications.\\
\noindent Our system gives results with guarantees, therefore, our interoperability testing surpasses the other interoperability techniques that have been offered for CAD model verification and validation in literature.
Ideally, a fully-automated system should include a single build environment to query the systems and run \textit{DTest}, and allow investigating the tolerance levels that would allow the systems to interoperate with respect to the given properties if a negative interoperability result is reached within the specified tolerance. However, the technical challenges and the short project duration hindered the progress towards building this ideal system.\\
\noindent In future research, we plan to update the template files with additional query responses and properties, and experiment with different CAD systems and file formats. We would also like to integrate the existing commercial interoperability checks for the verification of interoperability and translation validation into our automated system approach.
  
\appendix
%\newpage 
\section{Template Files for the Coil Model}
%\newpage 
\begin{figure}[tbh!]
    \centering
    \includegraphics[width=0.5\textwidth]{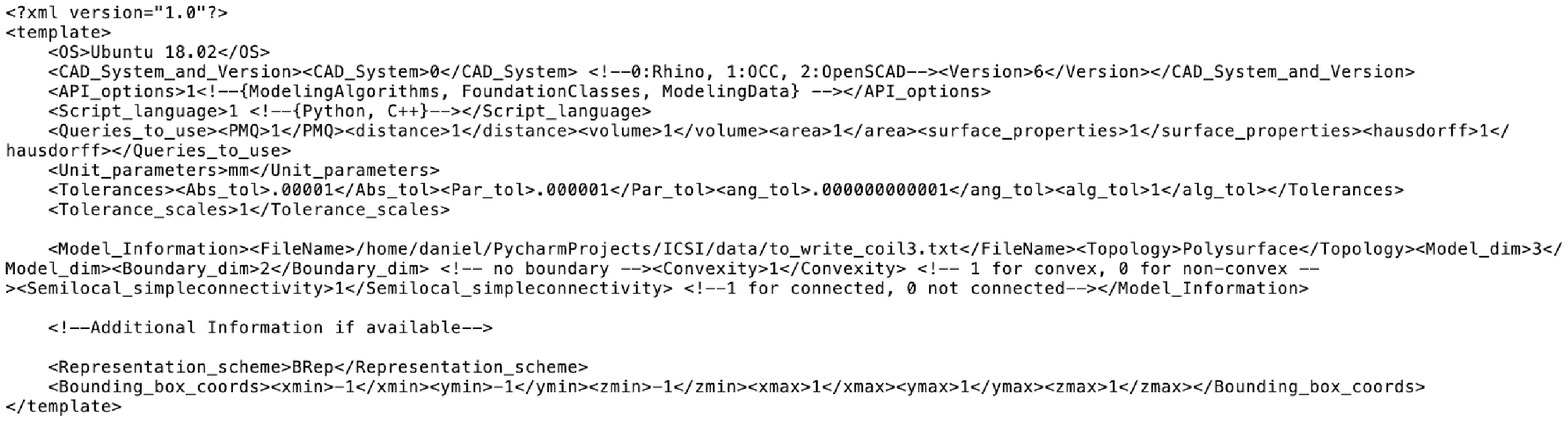}
    \caption{Template file for the Rhino model}
    \label{fig:rc}
\end{figure}

\begin{figure}[tbh!]
    \centering
    \includegraphics[width=0.5\textwidth]{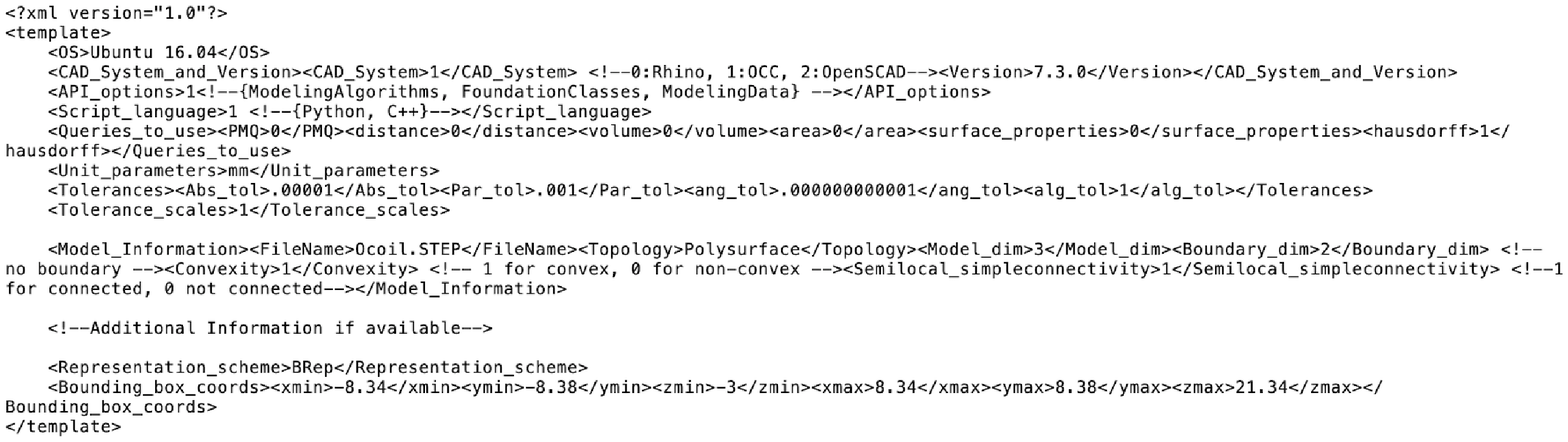}
    \caption{Template file for the OpenCASCADE model}
    \label{fig:oc}
\end{figure}
\section{Software Components}
We used the following software and libraries in this research:
\begin{itemize}
    \item OpenCASCADE \cite{occ}: An object-oriented C++ class library designed for fast production of advanced domain-specific CAD/CAM/CAE applications.
    \item PythonOCC \cite{occ}: A 3D CAD/CAE/PLM development framework for Python. It includes features such as advanced topological and geometrical operations, data exchange (STEP, IGES, STL import/export), GUI based visualization (wx, Qt), jupyter notebook rendering.
    \item OpenCASCADE Community Edition (OCE) \cite{oce} :A C++ 3D modeling library. It can be utilized to develop CAD/CAM softwares such as FreeCad or IfcOpenShell. It aims to gather patches/changes/improvements from the OCC community. 
    \item Rhinoceros 5 \cite{rhino} : A 3D computer graphics and CAD application software developed by Robert McNeel \& Associates. Rhinoceros geometry is based on the NURBS model, which focuses on building mathematically precise representation of curves and freeform surfaces in computer graphics (as opposed to polygon mesh-based applications).
    \item Structural Bioinformatics Library (SBL) \cite{sbl}: A template C++/Python library for solving structural biology problems. It provides programs (executables) for end-users and a rich framework to develop new applications.
\end{itemize}
\noindent The code for the automated system presented here is not publicly available due to the ongoing improvements but could be provided on reasonable request. 
%\end{verbatim}
%\newpage
\section*{Acknowledgments}
\noindent This research is funded by the DARPA contracts HR00111620042 and HR0011623402. We would like to thank Vadim Shapiro from the University of Wisconsin-Madison and ICSI, Berkeley, USA, for his helpful comments and support.

%\section{Front matter}
%
%The author names and affiliations could be formatted in two ways:
%\begin{enumerate}[(1)]
%\item Group the authors per affiliation.
%\item Use footnotes to indicate the affiliations.
%\end{enumerate}
%See the front matter of this document for examples. You are recommended to conform your choice to the journal you are submitting to.

%\section{Bibliography styles}
%
%There are various bibliography styles available. You can select the style of your choice in the preamble of this document. These styles are Elsevier styles based on standard styles like Harvard and Vancouver. Please use Bib\TeX\ to generate your bibliography and include DOIs whenever available.
%
%Here are two sample references: \cite{Feynman1963118,Dirac1953888}.

%\section*{References}

%\bibliography{mybibfile}

\end{document}